\documentclass[conference]{IEEEtran}
\IEEEoverridecommandlockouts
\usepackage{cite}
\usepackage{amsmath,amssymb,amsfonts}
\usepackage{algorithmic}
\usepackage{graphicx}
\usepackage{textcomp}
\usepackage[hidelinks]{hyperref}
\usepackage{xcolor}
\usepackage{array}
\usepackage{booktabs}
\def\BibTeX{{\rm B\kern-.05em{\sc i\kern-.025em b}\kern-.08em
    T\kern-.1667em\lower.7ex\hbox{E}\kern-.125emX}}

\begin{document}

\title{TuneBench: Benchmarking AI Agents for Power System Dynamic Model Calibration\\
}

\author{\IEEEauthorblockN{Andrea Pomarico}
\IEEEauthorblockA{\textit{Department of Energy} \\
\textit{Politecnico di Milano}\\
Milan, Italy \\
andrea.pomarico@polimi.it}
\and
\IEEEauthorblockN{Salvatore Tessitore}
\IEEEauthorblockA{\textit{Department of System Engineering} \\
\textit{Terna Rete Italia S.p.A.}\\
Rome, Italy \\
salvatore.tessitore@terna.it}
\and
\IEEEauthorblockN{Giorgio Giannuzzi}
\IEEEauthorblockA{\textit{Department of System Engineering} \\
\textit{Terna Rete Italia S.p.A.}\\
Rome, Italy \\
giorgio.giannuzzi@terna.it}
\and
\IEEEauthorblockN{Alberto Berizzi}
\IEEEauthorblockA{\textit{Department of Energy} \\
\textit{Politecnico di Milano}\\
Milan, Italy \\
alberto.berizzi@polimi.it}
}

\maketitle
\begin{abstract}
Large Language Model (LLM)-based agents can automate complex engineering workflows by interacting with tools, interpreting feedback, and making iterative decisions. This paper introduces \textit{TuneBench}, a benchmark for evaluating Agentic AI in power system dynamic model calibration, where controller parameters are adjusted to reproduce target dynamic responses. The benchmark reflects a common Transmission System Operator (TSO) practice, in which equivalent models used for Dynamic Security Assessment (DSA) are calibrated against measurements from grid-code compliance tests when proprietary models are unavailable. Two cases of increasing complexity are considered: the governor and Automatic Voltage Regulator (AVR) of a synchronous generator and the control system of a Battery Energy Storage System (BESS), both tuned against active-power and voltage step tests within a budget of 15 simulations. LLM-based agents are compared with random search and Genetic Algorithm (GA) baselines over 10 independent runs. Only the most advanced models, Opus 5.5 and GPT-6, accurately reproduce all target trajectories, with median $R^2$ above 0.995 in every test and tracking errors up to two orders of magnitude lower than the baselines. The agent logs show that these agents autonomously extract dynamic performance indices, such as settling time and damping, and relate them to specific control parameters, decomposing the tuning problem into physically interpretable sub-problems. The benchmark is publicly available at \href{https://github.com/Power-Agent/PowerAgentBench/tree/main/benchmarks/dynamic/level3}{https://github.com/Power-Agent/PowerAgentBench/tree/main/benchmarks/dynamic/level3}.
\end{abstract}


\section{Introduction}

In recent years, Transmission System Operators (TSOs) have increasingly integrated Dynamic Security Assessment (DSA) into Energy Management Systems (EMSs) to complement steady-state contingency analysis with time-domain simulations \cite{pannell2018machine}. While power flow analysis assesses the feasibility of post-contingency operating conditions, dynamic simulations capture the transient evolution of key variables, including frequency, voltage, and active and reactive power. The growing penetration of inverter-based resources (IBRs) further increases the importance of modeling fast control dynamics \cite{nerc2021odessa}. Although electromagnetic transient (EMT) simulations provide greater detail, their computational requirements limit their application to large-scale operational studies. Consequently, DSA predominantly relies on phasor-domain (RMS) simulations, balancing accuracy and computational efficiency \cite{ren2022online}. The reliability of these analyses depends strongly on the accuracy of the underlying dynamic models \cite{pinheiro2025evaluation}. Since proprietary models and control parameters are not always disclosed to TSOs, equivalent models often need to be developed and calibrated using measurements from grid-code compliance and commissioning tests, including active-power, reactive-power, and voltage-control tests. Model parameters are then iteratively adjusted to minimize the mismatch between simulated and measured trajectories, either manually or through optimization techniques such as genetic algorithms~\cite{stefopoulos2005genetic} and Kalman filtering~\cite{huang2017calibrating}.
Despite existing calibration approaches, this process still relies heavily on experienced engineers to identify relevant parameters, interpret their effects, and refine the model.

Recent advances in Agentic AI have enabled language models to execute code, interact with external tools, and adapt their decisions based on feedback, making them promising for multi-step power engineering workflows. Previous studies have investigated LLMs for electricity-market bidding~\cite{zou2025large}, unit commitment and dispatch~\cite{yang2025large}, optimal power flow~\cite{li2025optdispro,bernier2025powergraph}, and grid control~\cite{saha2025dragent}, alongside broader power system applications~\cite{jia2025enhancing,xie2026crucial}. However, many approaches address individual tasks rather than complete engineering workflows. More recently, agentic frameworks have been developed for distribution grid analysis~\cite{badmus2027powerchain} and complex power system problems~\cite{chen2025x}. This growing interest is also reflected by the \emph{PowerAgent} community~\cite{zhang2025poweragent} and its benchmarks, \emph{PowerAgentBench-SS}~\cite{mylonas2026poweragentbenchss} and \emph{PowerAgentBench-Dyn}~\cite{zhang2026poweragentbenchdyn}, targeting steady-state and dynamic studies, respectively. Similarly, \emph{RestoreBench} addresses power flow non-convergence recovery \cite{mansutti2026restorebench}.

This paper introduces \textit{TuneBench}, a benchmark for evaluating Agentic AI in power system dynamic model tuning. It assesses agents based on different LLMs in tasks requiring engineering knowledge, interpretation of dynamic responses, and iterative interaction with simulation tools. Unlike single-step approaches, agents must autonomously select and adjust model parameters based on simulation feedback. The main contributions of this paper are:
\begin{itemize}
    \item A general evaluation framework for agent-based dynamic studies, defining observations, actions, simulator feedback, constraints, and performance metrics.
    \item A dynamic model tuning benchmark inspired by practical TSO workflows, requiring agents to identify relevant control parameters and calibrate models against reference dynamic responses.
    \item A comparative evaluation of different LLM-based agents against random search and Genetic Algorithm (GA)-based optimization, considering tuning accuracy and computational requirements.
\end{itemize}
\section{TuneBench Framework}

We formulate each benchmark task as an interactive environment
\begin{equation}
\mathcal{T} = (\mathcal{O}, \mathcal{A}, \mathcal{E}, \mathcal{C}, \mathcal{G}, \mathcal{M}),
\end{equation}
where $\mathcal{O}$ represents the observation space, $\mathcal{A}$ the action space, $\mathcal{E}$ the simulation environment, $\mathcal{C}$ the user-defined and engineering constraints, $\mathcal{G}$ the hidden ground truth or evaluation mechanism, and $\mathcal{M}$ an optional semantic memory component. At each step $t$, the agent receives an observation $o_t \in \mathcal{O}$, selects an action $a_t \in \mathcal{A}$, and obtains a new observation $o_{t+1}$ through simulator or tool feedback. This process continues until a final solution is produced or the budget is exhausted.

The parameter search space is intentionally unrestricted, allowing agents to autonomously identify and modify any accessible model parameter based on its physical meaning and influence on the dynamic response. Since the benchmark employs models documented in the literature, parameter descriptions support the agents in identifying relationships between control settings and electrical variables, such as active power and voltage. Thus, parameter selection becomes part of the evaluated reasoning process rather than being predefined. Nevertheless, agent interactions remain subject to engineering constraints specified in a machine-readable file, including permitted event modifications, simulation budgets, protected files, and reporting requirements. This approach preserves agent autonomy while ensuring controlled and reproducible evaluations.

\begin{figure}[h]
\centering
\includegraphics[width=\columnwidth]{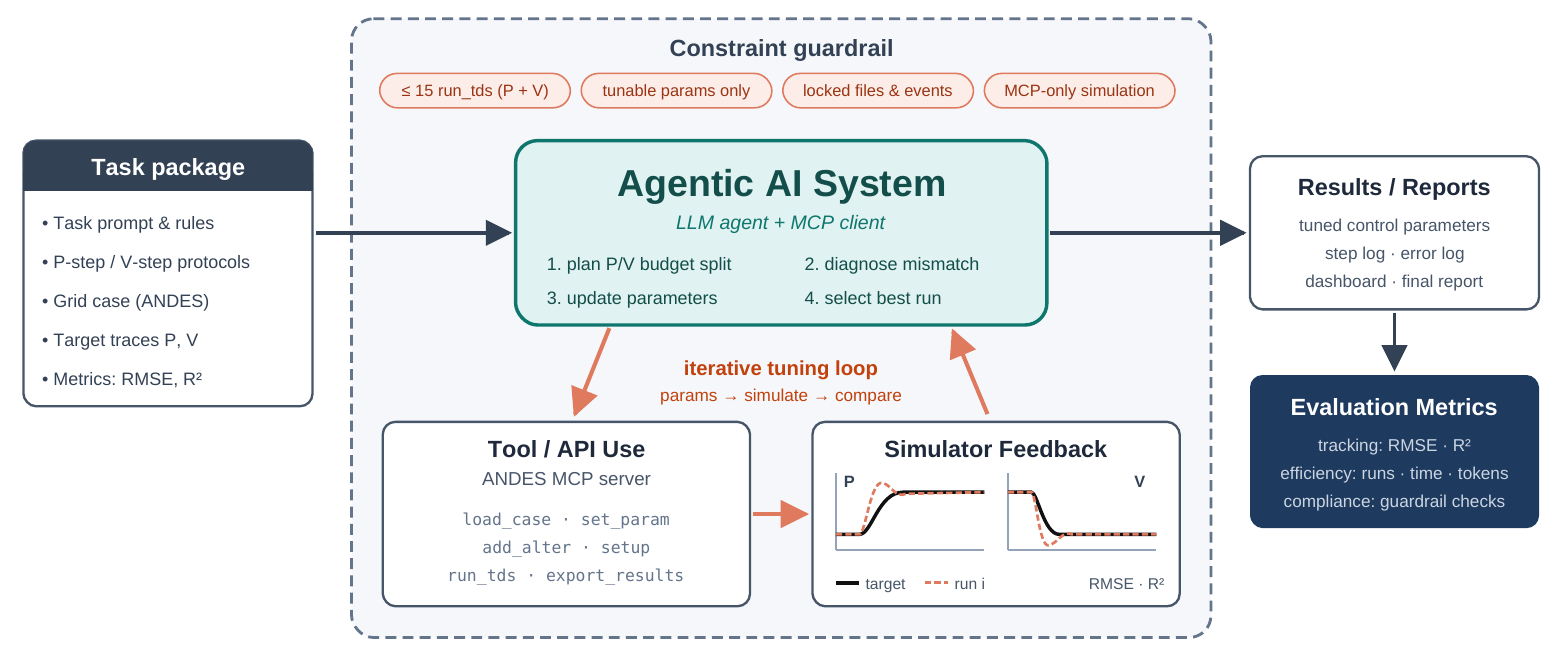}
\caption{Overview of the proposed TuneBench framework.}
\label{fig:framework_overview}
\end{figure}

As shown in Fig.~\ref{fig:framework_overview}, the framework follows four principles:
\begin{enumerate}
    \setlength{\itemsep}{0pt}
    \setlength{\parskip}{0pt}
    \setlength{\parsep}{0pt}
    \item \textit{Tool-grounded evaluation}: Agents interact with simulation and analysis tools, grounding their decisions in executable engineering workflows.
    \item \textit{Multi-round decision making}: Tasks require iterative actions and adaptation to simulation feedback rather than single-step responses or predefined scripts.
    \item \textit{Engineering constraint guardrails}: Agent actions must comply with simulation budgets, protected-file restrictions, and user-defined requirements.
    \item \textit{Outcome-oriented evaluation}: Performance is measured through engineering-relevant outcomes, including diagnostic accuracy, waveform agreement, and runtime.
\end{enumerate}

\subsection{Task Description}

Power system dynamic studies investigate the time-domain response to disturbances such as faults, equipment outages, frequency events, and control actions. Applications include transient and voltage stability, frequency response, oscillation damping, dynamic model validation, and fault ride-through assessment. These studies are commonly performed using simulation tools such as PSS/E, DIgSILENT PowerFactory, PSCAD, and ANDES~\cite{cui2018andes}.

\textit{TuneBench} addresses dynamic model calibration against target trajectories, reproducing a practical TSO workflow in which models used for DSA are tuned using measurements from grid-code compliance tests. Traditionally, experienced engineers iteratively adjust control parameters to match simulated and measured responses. In the proposed benchmark, this process is performed by an AI agent that autonomously configures and runs simulations, retrieves and stores results, compares simulated and target trajectories, and selects parameter updates based on the observed deviations. The procedure continues until satisfactory agreement is achieved or the simulation budget is exhausted.

\subsection{Evaluation Metrics}

To assess tuning accuracy, \textit{TuneBench} adopts two complementary metrics: the Root Mean Square Error (RMSE), which measures the deviation between simulated and target trajectories, and the coefficient of determination ($R^2$), which evaluates how well the simulation reproduces the variability of the target response.

Given $N$ samples of the target trajectory $y_i$ and simulated response $\hat{y}_i$, RMSE is defined as
\begin{equation}
    \mathrm{RMSE} =
    \sqrt{\frac{1}{N}\sum_{i=1}^{N}
    \left(y_i-\hat{y}_i\right)^2},
\end{equation}
where lower values indicate better agreement. The coefficient of determination is given by
\begin{equation}
    R^2 =
    1-
    \frac{\sum_{i=1}^{N}\left(y_i-\hat{y}_i\right)^2}
    {\sum_{i=1}^{N}\left(y_i-\bar{y}\right)^2},
\end{equation}
where $\bar{y}$ is the mean target value. Values of $R^2$ approaching one indicate a closer reproduction of the target response. Together, these metrics characterize both fitting error and agreement with the observed dynamics.

As summarized in Table~\ref{tab:task_config_dmqr}, the open-source simulator ANDES is integrated through dedicated tool functions. The agent executes dynamic simulations, retrieves waveforms and diagnostics, evaluates the mismatch using RMSE and $R^2$, and updates the relevant model parameters. This iterative process continues until the target response is adequately reproduced or the maximum number of simulations is reached.

\begin{table}[t]
\caption{Configuration of the \textit{TuneBench} benchmark.}
\label{tab:task_config_dmqr}
\centering
\scriptsize
\setlength{\tabcolsep}{3pt}
\renewcommand{\arraystretch}{1.1}
\begin{tabular}{@{}>{\raggedright\arraybackslash}p{0.22\columnwidth}
                 >{\raggedright\arraybackslash}p{0.72\columnwidth}@{}}
\toprule
\textbf{Item} & \textbf{Description} \\
\midrule
Simulator & ANDES. \\
Inputs & Initial dynamic model and target trajectories. \\
Agent actions & Configure and run simulations, analyze dynamic responses, and modify model parameters. \\
Parameter space & All accessible dynamic model parameters. \\
Simulation budget & Maximum of 15 simulations per benchmark instance. \\
Metrics & RMSE and $R^2$. \\
\bottomrule
\end{tabular}
\end{table}
\section{Numerical Results}

Both benchmark cases are built in ANDES. The device under test is connected to an infinite bus through a line and initially operates at $P_0=0.5$~pu and $V_0\approx1$~pu. Grid codes prescribe a wide range of compliance tests for generating units; among them, two of the most common ones are selected in this work to define the target behavior:
\begin{itemize}
    \item \textit{Test~P (active-power control):} the active-power reference is stepped from 0.5 to 0.7~pu at $t=5$~s, back to 0.5~pu at 35~s, down to 0.2~pu at 65~s and back to 0.5~pu at 95~s.
    \item \textit{Test~V (voltage control):} the voltage reference is stepped from its initial value to 0.97~pu at $t=5$~s.
\end{itemize}
The target parameter sets are not meant to give an optimal or grid-code-compliant dynamic response. They are designed to test the reasoning of the agents. If the default parameters of the standard IEEE and WECC models were used as targets, the LLMs could recall them from their prior knowledge of the literature and reach the solution without any real interaction with the simulator. The target parameters were therefore deliberately moved away from the standard values, even at the cost of a sub-optimal response, so that the agent must infer them from the observed mismatch and from its understanding of how each parameter affects the dynamic response.

The two tests are tuned sequentially. The 15-simulation budget covers both tests, and the agent decides how to split it between them. Finally, two verification runs, not counted in the budget, simulate both tests with the complete tuned parameter set to check the interaction between the two control loops. GA and random search are given a budget of 150 simulations. Each method is executed over 10 independent runs to account for its stochastic behavior. Agent setup, prompts, and baseline configurations are available in the repository.\footnote{The benchmark is publicly available at \href{https://github.com/Power-Agent/PowerAgentBench/tree/main/benchmarks/dynamic/level3}{\textcolor{blue}{https://github.com/Power-Agent/PowerAgentBench/tree/main/benchmarks/dynamic/level3}}}

\subsection{Synchronous Generator Tuning}

\begin{figure*}[t]
    \centering
    \includegraphics[width=.9\linewidth]{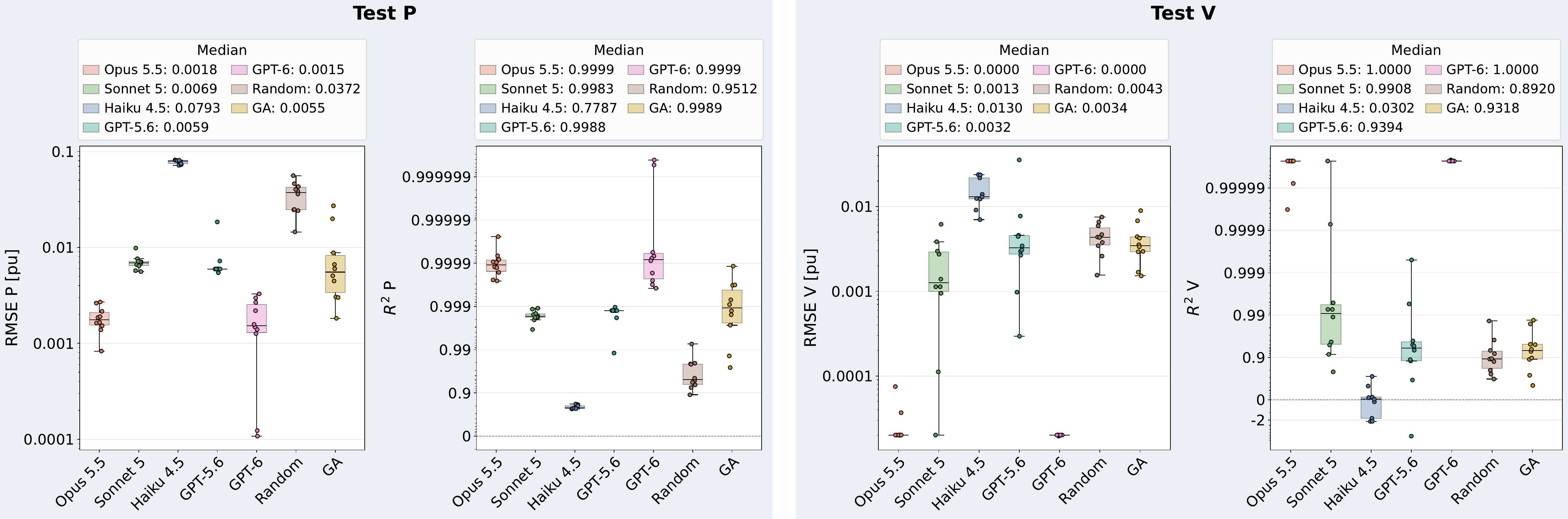}
    \caption{Generator tuning: distribution of RMSE and $R^2$ over 10 runs for the active-power (Test~P) and voltage (Test~V) control tests.}
    \label{fig:sg_boxplot}
\end{figure*}

\begin{figure}[t]
    \centering
    \includegraphics[width=1\linewidth]{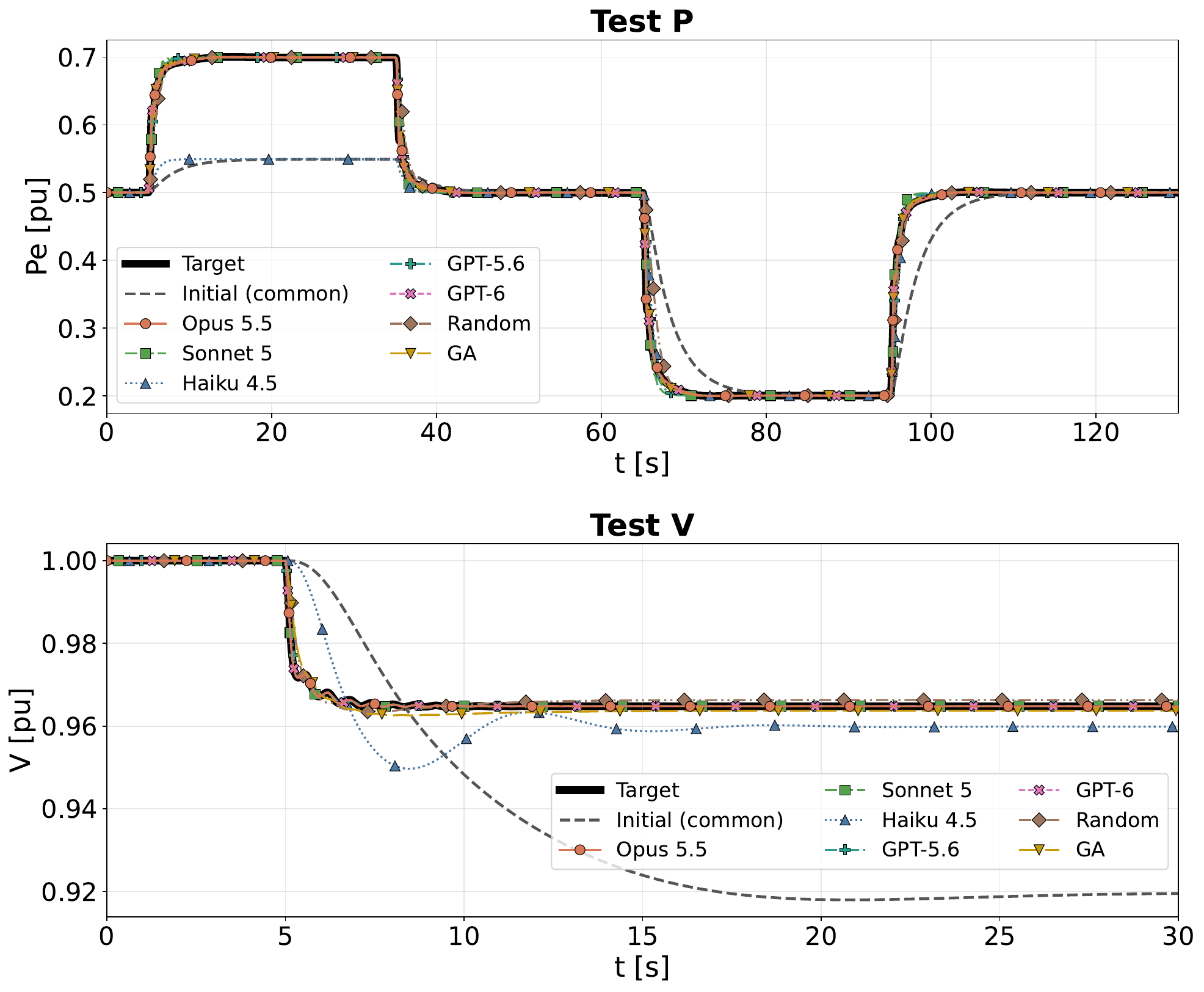}
    \caption{Generator tuning: best response obtained by each method compared with the target and the initial model for Test~P (top) and Test~V (bottom).}
    \label{fig:sg_best}
\end{figure}

The first case involves the tuning of the governor (IEEEG1) and AVR (ESST1A) of a synchronous generator. Fig.~\ref{fig:sg_boxplot} reports the distribution of RMSE and $R^2$ over the 10 runs, while Fig.~\ref{fig:sg_best} compares the best run of each method with the target trajectories. In Test~P, the initial model already reaches the active-power reference, but its response is slower than the target, especially during the downward steps. This makes Test~P the easiest task of the benchmark. Sonnet~5, GPT-5.6 and GA all achieve $R^2$ above 0.998, with median RMSE between $5.5\times10^{-3}$ and $6.9\times10^{-3}$~pu, while Opus~5.5 and GPT-6 obtain the lowest errors ($1.8\times10^{-3}$ and $1.5\times10^{-3}$~pu, $R^2=0.9999$). Random search ($R^2=0.951$) and Haiku~4.5 ($R^2=0.779$) perform worst, and the best Haiku~4.5 run settles at about 0.55~pu instead of 0.7~pu, i.e., it is even worse than the initial model.

Test~V is considerably more selective. The initial model shows a much slower voltage response and settles at about 0.92~pu instead of 0.965~pu, so both the dynamics and the steady-state behavior of the AVR must be corrected. Opus~5.5 and GPT-6 recover the target almost exactly in practically all runs, with median RMSE below $10^{-5}$~pu and $R^2=1$. Sonnet~5 reaches a median $R^2$ of 0.991, but with a large dispersion: some of its runs find the exact solution while others do not. GPT-5.6, GA and random search are limited to $R^2$ between 0.89 and 0.94 (RMSE of $3.2$--$4.3\times10^{-3}$~pu), whereas Haiku~4.5 fails the task ($R^2=0.03$).

\subsection{BESS Tuning}

\begin{figure*}[t]
    \centering
    \includegraphics[width=.9\linewidth]{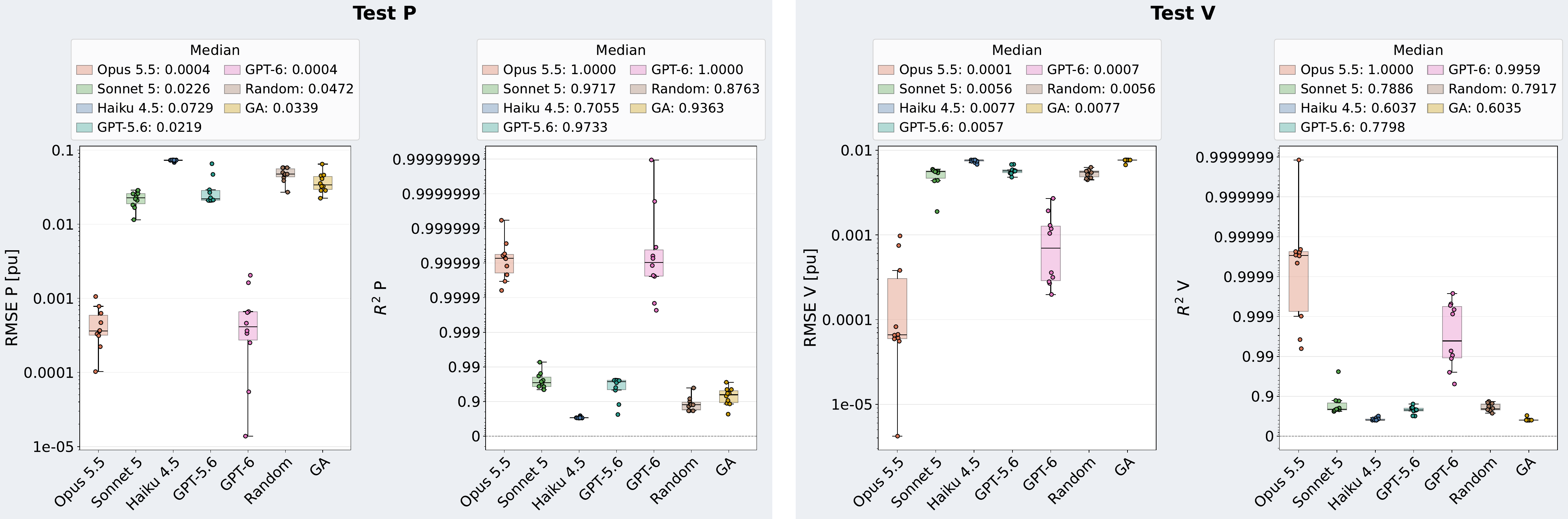}
    \caption{BESS tuning: distribution of RMSE and $R^2$ over 10 runs for the active-power (Test~P) and voltage (Test~V) control tests.}
    \label{fig:boxplot_test_summary_row}
\end{figure*}

\begin{figure}[t]
    \centering
    \includegraphics[width=1\linewidth]{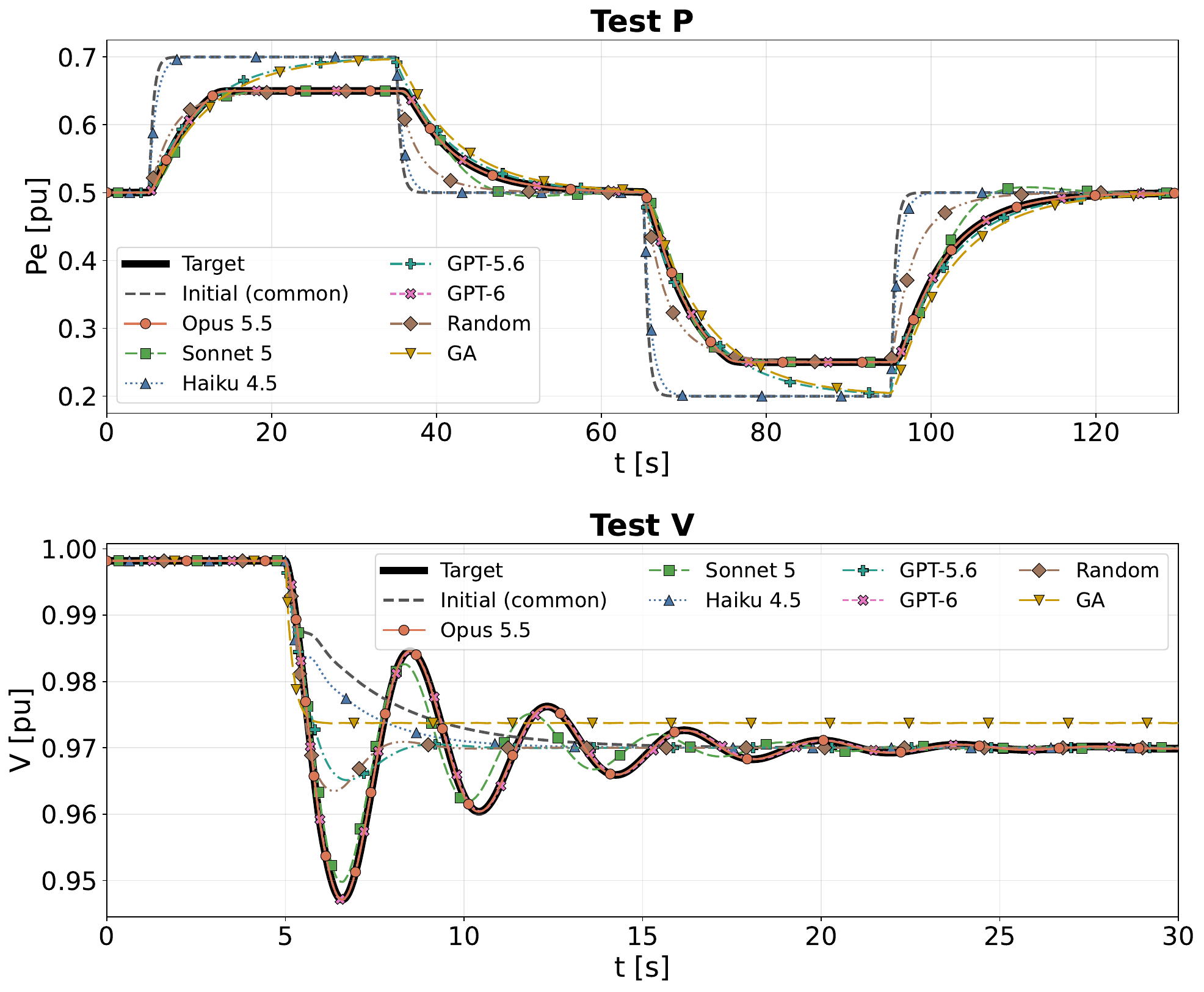}
    \caption{BESS tuning: best response obtained by each method compared with the target and the initial model for Test~P (top) and Test~V (bottom).}
    \label{fig:best_models_vs_target}
\end{figure}

The BESS is represented by the WECC generic models REGCA1 (converter interface), REECA1 (electrical control) and REPCA1 (plant-level control). In both tests, the reference steps are applied to the plant-level controller REPCA1. Compared with the generator, this case is considerably more challenging and clearly separates the methods into two groups (Figs.~\ref{fig:boxplot_test_summary_row} and~\ref{fig:best_models_vs_target}).

In Test~P, the initial model tracks the active-power reference almost instantaneously, whereas the target exhibits slower ramps and does not reach the reference at steady state (about 0.65 and 0.25~pu instead of 0.7 and 0.2~pu). Opus~5.5 and GPT-6 are the only agents that consistently recover both features, achieving median RMSE values of $3.6\times10^{-4}$ and $4.1\times10^{-4}$~pu, i.e., about two orders of magnitude lower than random search ($4.7\times10^{-2}$~pu) and GA ($3.4\times10^{-2}$~pu), with median $R^2 \simeq 1$. GA reproduces the slower ramps but still settles at the reference value, and Sonnet~5 and GPT-5.6 only marginally improve on the baselines ($R^2\approx0.97$). Haiku~4.5 performs worse than random search ($R^2=0.705$), since its best solution remains close to the initial model.

In Test~V, the initial model shows an overdamped voltage response, whereas the target exhibits a lightly damped oscillation that the agent must reproduce in both frequency and damping. Only Opus~5.5 and GPT-6 succeed. Opus~5.5 achieves the best accuracy, with a median RMSE of $6.6\times10^{-5}$~pu and $R^2=0.99997$, roughly one order of magnitude better than GPT-6 ($7.0\times10^{-4}$~pu, $R^2=0.996$). All the remaining methods reach $R^2$ values below $0.8$. They converge to overdamped responses or, in the case of GA, to a steady-state voltage that does not match the reference.

\subsection{Computational Cost}

The wall-clock runtimes are summarized in Table~\ref{tab:runtime}. A substantial portion of the execution time is associated with reasoning between simulations. Opus~5.5 requires a median of 22.1~min for the generator and 67.5~min for the BESS, with considerable variability across runs (29--114~min), suggesting that reasoning effort increases with task complexity. GPT-6 provides the most favorable balance between accuracy and computational time, achieving performance comparable to Opus~5.5 with median runtimes of 17.6 and 21.7~min for the generator and BESS, respectively. Conversely, faster models such as Haiku~4.5 and GPT-5.6 do not achieve comparable tuning accuracy.

\begin{table}[t]
\caption{Wall-clock runtime per run over 10 runs [min].}
\label{tab:runtime}
\centering
\scriptsize
\setlength{\tabcolsep}{3pt}
\renewcommand{\arraystretch}{1.1}
\begin{tabular}{@{}>{\raggedright\arraybackslash}p{0.19\columnwidth}
                 *{6}{>{\centering\arraybackslash}p{0.11\columnwidth}}@{}}
\toprule
 & \multicolumn{3}{c}{\textbf{Generator}} & \multicolumn{3}{c}{\textbf{BESS}} \\
\cmidrule(lr){2-4}\cmidrule(l){5-7}
\textbf{Method} & \textbf{Median} & \textbf{Min} & \textbf{Max} & \textbf{Median} & \textbf{Min} & \textbf{Max} \\
\midrule
Opus 5.5   & 22.1 & 13.7 & 42.9 & 67.5 & 29.2 & 114.0 \\
Sonnet 5   & 21.2 & 16.3 & 27.7 & 30.3 & 26.8 & 41.5 \\
Haiku 4.5  & 11.9 & 3.1  & 45.0 & 9.3  & 5.4  & 20.1 \\
GPT-5.6    & 14.6 & 9.5  & 18.4 & 14.8 & 10.9 & 27.3 \\
GPT-6      & 17.6 & 8.7  & 24.1 & 21.7 & 14.2 & 34.9 \\
\bottomrule
\end{tabular}
\end{table}

\section{Discussion}

Only the most advanced models, Opus~5.5 and GPT-6, successfully complete the entire benchmark within the 15-simulation budget. The remaining approaches achieve competitive results mainly on the simplest task, generator Test~P, but struggle with more complex dynamic responses. Analysis of the agent logs provides insight into these differences. Without explicit instructions, the best-performing agents independently extract conventional dynamic performance indices, including settling time, ramp rate, oscillation frequency, and damping, from both simulated and target trajectories. By relating the observed deviations to the corresponding control parameters, they decompose tuning problem into smaller, physically interpretable sub-problems. This approach enables tracking errors up to two orders of magnitude lower than those obtained with GA and random search, which rely on a scalar error metric. Moreover, since the target parameters were deliberately shifted from their nominal values, the results suggest that the agents exploit knowledge of system dynamics rather than simply reproducing default settings. The resulting parameter adjustments are also supported by measurable performance indices, allowing TSO engineers to review and validate the tuning process.

\section{Conclusion}

This paper presented \textit{TuneBench}, a benchmark for assessing Agentic AI in power system dynamic model calibration, reflecting the practical need of TSOs to reproduce measured device responses when proprietary models are unavailable. Two tuning tasks were considered: the governor and AVR of a synchronous generator and the control system of a BESS. Among the evaluated approaches, Opus~5.5 and GPT-6 were the only models capable of accurately reproducing all target trajectories within 15 simulations, outperforming both the other LLM agents (Haiku~4.5, Sonnet~5, and GPT-5.6) and conventional random-search and GA baselines, with tracking errors reduced by up to two orders of magnitude. Future work will extend the benchmark from parameter calibration to dynamic model identification, where agents must determine not only the parameter values but also the underlying control structure.

\bibliography{biblio_agenticAI}

@article{jia2025enhancing,
  title={{Enhancing LLMs for power system simulations: A feedback-driven multi-agent framework}},
  author={Jia, Mengshuo and Cui, Zeyu and Hug, Gabriela},
  journal={IEEE Transactions on Smart Grid},
  year={2025},
  publisher={IEEE}
}

@inproceedings{zou2025large,
  title={{A Large Language Model-Based Agent for Automated Bidding Strategy Generation in Electricity Markets}},
  author={Zou, Ruixi and others},
  booktitle={2025 IEEE International Conference on Power and Integrated Energy Systems (ICPIES)},
  pages={475--480},
  year={2025},
  organization={IEEE}
}

@article{yang2025large,
  title={Large language model powered automated modeling and optimization of active distribution network dispatch problems},
  author={Yang, Xu and others},
  journal={IEEE Transactions on Smart Grid},
  year={2025},
  publisher={IEEE}
}

@article{li2025optdispro,
  title={{OptDisPro: LLM-based multi-agent framework for flexibly adapting heuristic optimal DisFlow}},
  author={Li, Zhengbo and others},
  journal={IEEE Transactions on Smart Grid},
  year={2025},
  publisher={IEEE}
}

@article{bernier2025powergraph,
  title={{PowerGraph-LLM: Novel power grid graph embedding and optimization with large language models}},
  author={Bernier, Fabien and others},
  journal={IEEE Transactions on Power Systems},
  year={2025},
  publisher={IEEE}
}

@inproceedings{saha2025dragent,
  title={DrAgent: An agentic approach to fault analysis in power grids using large language models},
  author={Saha, Barun Kumar and Aarthi, V and Naidu, OD},
  booktitle={2025 International Conference on Artificial Intelligence in Information and Communication (ICAIIC)},
  pages={0938--0945},
  year={2025},
  organization={IEEE}
}

@article{badmus2027powerchain,
  title={{Powerchain: A verifiable agentic ai system for automating distribution grid analyses}},
  author={Badmus, Emmanuel O and others},
  journal={Electric Power Systems Research},
  volume={262},
  pages={113555},
  year={2027},
  publisher={Elsevier}
}

@article{chen2025x,
  title={{X-GridAgent: An LLM-powered agentic AI system for assisting power grid analysis}},
  author={Chen, Xin and others},
  journal={arXiv preprint arXiv:2512.20789},
  year={2025}
}

@article{zhang2025poweragent,
  title={PowerAgent: A road map toward agentic intelligence in power systems: Foundation model, model context protocol, and workflow},
  author={Zhang, Qian and Xie, Le},
  journal={IEEE Power and Energy Magazine},
  volume={23},
  number={5},
  pages={93--101},
  year={2025},
  publisher={IEEE}
}

@article{xie2026crucial,
  title={{Crucial Role of Foundation Models in Enhancing the Interaction of AI and Power Systems: Achieving Integrated Frameworks}},
  author={Xie, Le and others},
  journal={IEEE Energy Sustainability Magazine},
  volume={2},
  number={1},
  pages={22--28},
  year={2026},
  publisher={IEEE}
}

@article{mylonas2026poweragentbenchss,
  title        = {{PowerAgentBench-SS: A Benchmark for Agentic AI in Power System Steady-State Studies}},
  author       = {Mylonas, Costas and others},
  year         = {2026},
  journal = {arXiv}
}

@article{zhang2026poweragentbenchdyn,
  title         = {{PowerAgentBench-Dyn: A Benchmark for Agentic AI in Power System Dynamic Studies}},
  author        = {Zhang, Qian and others},
  year         = {2026},
  journal = {arXiv}
}

@techreport{nerc2021odessa,
  author       = {{North American Electric Reliability Corporation} and {Texas Reliability Entity}},
  title        = {Odessa Disturbance Texas Events: May 9, 2021 and June 26, 2021},
  year         = {2021},
  month        = sep,
  institution  = {NERC},
}

@article{ren2022online,
  title={Online multi-fault power system dynamic security assessment driven by hybrid information of anticipated faults and pre-fault power flow},
  author={Ren, Junyu and others},
  journal={International Journal of Electrical Power \& Energy Systems},
  volume={136},
  pages={107651},
  year={2022},
  publisher={Elsevier}
}

@inproceedings{pannell2018machine,
  title={Machine learning approach to solving the transient stability assessment problem},
  author={Pannell, Zachary and Ramachandran, Bhuvaneswari and Snider, Dallas},
  booktitle={2018 IEEE Texas power and energy conference (TPEC)},
  pages={1--6},
  year={2018},
  organization={IEEE}
}

@article{pinheiro2025evaluation,
  title={{Evaluation of the impact of IBR on the frequency dynamics in the Brazilian power system}},
  author={Pinheiro, Bruno and others},
  journal={Sustainable Energy, Grids and Networks},
  pages={101827},
  year={2025},
  publisher={Elsevier}
}

@inproceedings{cui2018andes,
  title={{ANDES: A Python-based cyber-physical power system simulation tool}},
  author={Cui, Hantao and Li, Fangxing},
  booktitle={2018 North American Power Symposium (NAPS)},
  pages={1--6},
  year={2018},
  organization={IEEE}
}

@article{huang2017calibrating,
  title={Calibrating parameters of power system stability models using advanced ensemble Kalman filter},
  author={Huang, Renke and others},
  journal={IEEE Transactions on Power Systems},
  volume={33},
  number={3},
  pages={2895--2905},
  year={2017},
  publisher={IEEE}
}

@inproceedings{stefopoulos2005genetic,
  title={A genetic algorithm solution to the governor-turbine dynamic model identification in multi-machine power systems},
  author={Stefopoulos, George K and others},
  booktitle={Proceedings of the 44th IEEE Conference on Decision and Control},
  pages={1288--1294},
  year={2005},
  organization={IEEE}
}

@article{mansutti2026restorebench,
  title={{RestoreBench: Can AI Agents Restore Power Flow Convergence?}},
  author={Mansutti, Riccardo and Pomarico, Andrea and Jakob, Robert and Zhang, Qian and Berizzi, Alberto and O'Sullivan, Kevin},
  journal={arXiv preprint arXiv:2609.00384},
  year={2026}
}
\bibliographystyle{IEEEtran}

\end{document}